\documentclass[%
 reprint,
 amsmath,amssymb,
 aps,
]{revtex4-2}

\usepackage{amsmath,amssymb,amsfonts}
\usepackage{algorithmic}
\usepackage{graphicx}
\usepackage{gensymb}
\usepackage{subcaption}

\usepackage{graphicx}
\usepackage{dcolumn}
\usepackage{bm}

\begin{document}

\preprint{APS/123-QED}

\title{Frequency Stability of Spin-Hall Nano-Oscillators with Realistic Grain Structure}

\author{Corrado Carlo Maria Capriata}
 \email{capriata@kth.se}
\author{Bengt Gunnar Malm}
 \affiliation{Division of Electronics and Embedded Systems, KTH Royal Institute of Technology, Stockholm, Sweden}
 
\author{Sheng Jiang}%
 \affiliation{Department of Physics, University of Gothenburg, Gothenburg, Sweden}%

\author{Johan \AA kerman}
\affiliation{Department of Physics, University of Gothenburg, Gothenburg, Sweden}
\affiliation{NanOsc AB, Kista 164 40, Sweden}%

\author{Mykola Dvornik}
\affiliation{NanOsc AB, Kista 164 40, Sweden}%


\date{\today}

\begin{abstract}
Nano-constriction spin-Hall nano-oscillators (NC-SHNOs) are one of the most promising alternatives among the microwave spintronics devices. They can provide highly coherent and widely tuneable microwave signals and can be fabricated at low temperature, which makes them compatible with back-end-of-the-line CMOS processing. For its applications, the frequency stability of each device is crucial, in particular for synchronization of oscillator arrays. In this work, we focus on the influence of a realistic grain structure on the SHNO frequency stability using both measurements as well as micromagnetic simulations. Grains in the thin ferromagnetic metal films can influence the output characteristic of the SHNO since the exchange coupling is reduced locally at the grain boundaries. This work provides a novel micromagnetic simulation method, for systematic investigation of frequency instability or variability from device-to-device. Experimentally, a device-to-device frequency variability of $\sim 270 MHz$ was found and in some extreme cases, a double-mode oscillation was observed in the high current operation range. This oscillation behavior was reproduced in simulations, where the inclusion of grains resulted in a frequency variability of $\sim 100 MHz$ and double-mode oscillations for some particular configurations. The double modes are consistent with a partial decoupling, or non-coherent operation, of two oscillating regions located at the nano-constriction edges.
\end{abstract}

\maketitle


\section{Introduction}
Considering the field of nano-oscillators \cite{Dieny2020}, the nano-constriction spin-Hall nano-oscillators (NC-SHNO) are one of the best candidates among the microwave spintronic devices. In fact, they are compatible with low-temperature, back-end-of-the-line, CMOS processing techniques \cite{chen2016spin, malm2019micromagnetic, Zahedinejad2018}. Both single\cite{Demidov2014,Dvornik2018, Mazraati2018, awad2020}, 1-dimensional \cite{awad2017long}, and 2-dimensional\cite{zahedinejad2020two} nano-constriction devices have been shown to provide highly coherent and widely tuneable microwave signals at room temperature. All SHNOs use the spin Hall effect to operate \cite{Demidov2012, Hirsch1999, zhang2000, Kato2004, Saitoh2006}, but different geometries, such as nano-wires, nano-gaps, and nano-constrictions, were proposed to optimize the device \cite{zholud2014microwave, Duan2014, durrenfeld201720}. More recently, the NC-SHNO was proposed as a solution as an easier fabrication process and lower driving current are required. During the years, different fabrication techniques were studied to optimise the processing of NC-SHNOs and different combinations of materials were tested as well. The most commonly used metal bi-layers are CoFeB/Pt, NiFe/Pt and NiFe/W \cite{Ranjbar2014, Mazraati2016}, however more recently even single layer devices were demonstrated \cite{Haidar2019}.\par
Mode coherence and frequency stability are crucial for future applications, utilizing multi-constriction NC-SHNO \cite{awad2017long} or NC-SHNO array configurations. \cite{zahedinejad2020two}. For this reason, it is of great interest to analyse the device-to-device output characteristics variability at conditions as close as possible to the real application use. In this work, we performed experimental measurements and micromagnetic simulations of NC-SHNOs, with an out-of-plane applied field at room temperature for devices with constriction widths ranging from 80 nm to 200 nm. The aim of the investigations was to examine frequency variability (device-to-device) and to analyze possible multi-mode behavior.\par
Our measurements highlight the presence of double-mode oscillations and high-current divergence with respect to ideal devices. These effects are found from mid-high currents ($I \sim $ 2.5 mA), which means that most of the devices show limited variability  in the usual working regime ($I \sim 1-2$ mA).\par
Multi-mode behaviour of microwave spintronic oscillators has been studied in some previous work. Regarding spin-torque nano-oscillators (STNO) both mode coexistence and mode hopping have been thoroughly explored \cite{dumas2013spin, iacocca2015mode, Iacocca2014, Sharma2014, Heinonen2013, Zhang2017, Eklund2014}. Here, two different coupling mechanism are identified: inter-mode interactions, and spinwave (magnon) scattering. Some previous work has addressed the coupling of multiple oscillation modes within single nano-gap SHNO devices \cite{chen2019, chen2019dynamical}. However, in those experiments, the devices were measured at low temperature and with an in-plane applied field. These studies did not relate multi-mode behavior and device variability to material properties or fabrication variability but rather to the interaction of possible frequency modes, supported by the device itself. Regarding SHNOs the most recent study is the one by Chen et al. \cite{chen2019dynamical}, where the authors link the presence, or co-existing condition, of two modes at high currents to their small spatial overlap.\par
In this study, we specifically address the local exchange coupling variations that are inherently present in a thin ferromagnetic film that exhibits a grain structure. This effect could act to partially break the mode phase and frequency coherence and allow for multiple simultaneous oscillation modes. \par
The rest of this article is structured as follows. First, both the methods and the experimental results on NC-SHNOs are introduced in Section II. The methods and results of the micromagnetic simulations are then presented in Section III. A more in-depth micromagnetic study regarding the double-mode coexistence conditions for our SHNO configurations is then presented in Section IV. Finally, Section V concludes this article.

\section{Measurements}

\subsection{Methods}
\begin{figure}[ht!]
\centering
\includegraphics[width=0.45\textwidth]{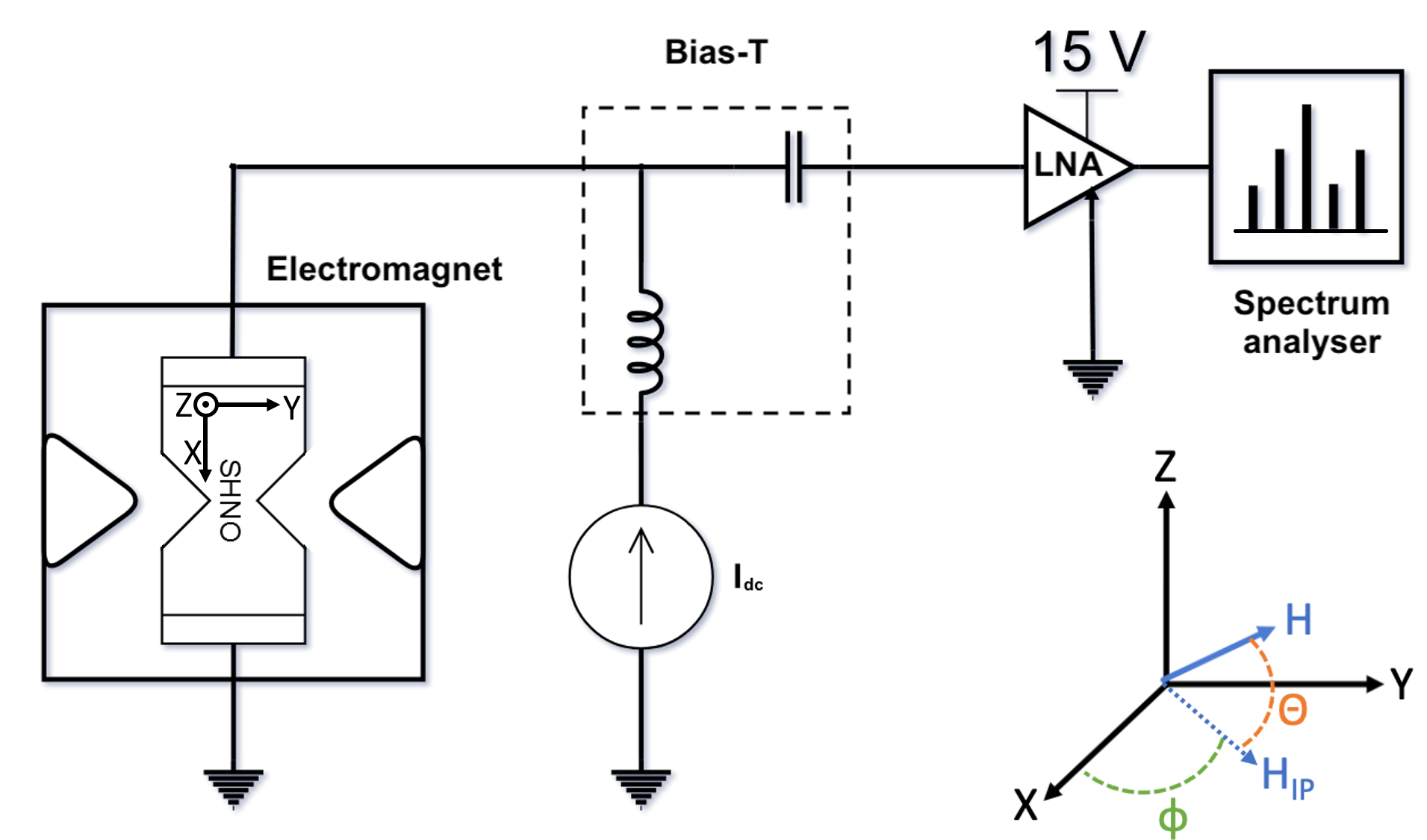}
\caption{Schematic of the measurement setup with coordinate system, where a magnetic field is applied at an IP-angle $\phi $ and an OOP-angle $\theta $.}
\label{fig:circuit}
\end{figure}

All measurements were performed on a single chip with a large number of SHNOs\cite{awad2020}, featuring different constriction widths ( from 80 - 200 nm), and in individual or multiple-oscillator configurations. This study focuses solely on the individual constriction devices. On each line of the chip several identical devices, patterned with the same width, were present and this was utilized in our design-of-experiments. The film structure of the SHNOs is composed of a 6 nm Pt and a 5 nm Py (Ni$_{80}$Fe$_{20}$) bilayer. After mounting the chip in the measurement setup shown in Figure \ref{fig:circuit}, the devices were manually probed one by one. A static magnetic field was applied at oblique direction with an in-plane (IP) angle $\phi $ and an out-of-plane (OOP) angle $\theta $ as shown in Figure \ref{fig:circuit}). The applied field was between 5 kOe and 6.5 kOe. A direct current $I_{dc}$ was applied to the device using a Keithley 6221 current source, the applied current ranged from  1 mA to 3.5 mA with 0.01 mA step. The lower current limit roughly corresponds to threshold current of the fundamental oscillating mode and the upper limit of sustained non-destructive device operation respectively. At even higher currents, devices were typically destroyed by combined Joule heating and contact electro-migration. These two effects are coupled since a temperature raise, induced by the Joule heating, would also act to accelerate the electo-migration. The output microwave signal of the device was first amplified by using a wideband, 38 dB gain, low-noise-amplifier (LNA) and then analysed by a Rohde\&Schwarz FSU-67 (20 Hz - 67 GHz) spectrum analyser. The sweep time was set to 50 ms (lower limit of the FSU-67). This short sweep time allows a quasi-real-time analysis of the frequency stability and double-mode presence. A bias-T was utilized to decouple the applied currents and output microwave signals.

\subsection{Results and Discussion}
\begin{figure}[ht!]
\centerline{\includegraphics[width=0.45\textwidth]{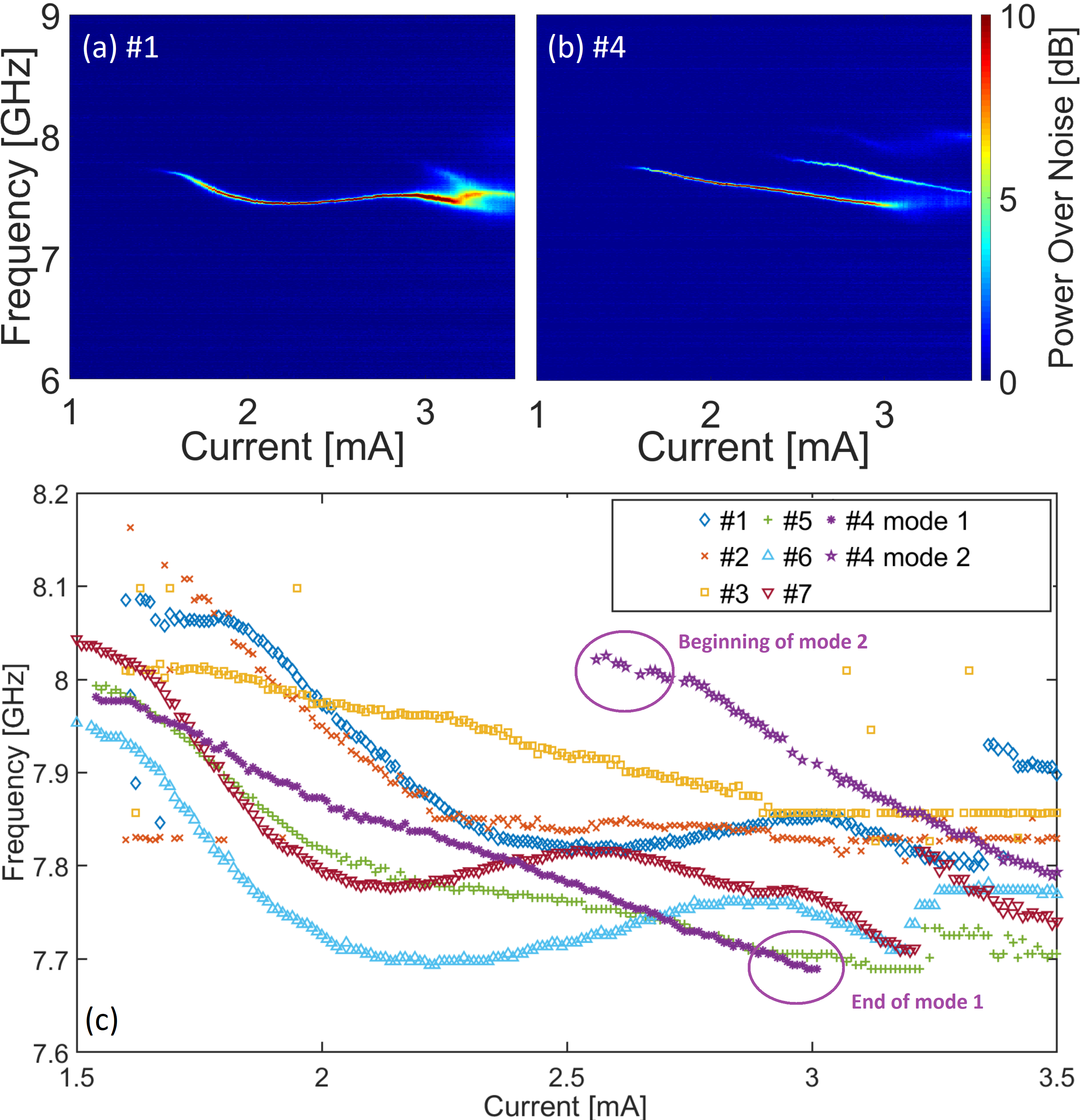}}
\caption{Representative PSDs for two different SHNOs with the same constriction  width (120 nm). Applied field 6 kOe at $\phi=20\degree $ and $\theta=80\degree$. (a) typical spectrum and (b) example of double modes; (c) Extracted frequencies for different devices with 120 nm width, highlighting a $\Delta f \approx 270 MHz$. Double-mode behaviour is typically observed above $\sim 3$ mA}
\label{fig:ideal_meas}
\end{figure}

From the mapping, we could find the typical behavior for many SHNOs. This behavior was to a large extent independent of the constriction width and therefore the width influence on the oscillating volume will not be discussed further. In the power spectral density (PSD), shown in Figure \ref{fig:ideal_meas}(a), one can observe an example of such behaviour. The oscillating frequency exhibits a clear output with narrow linewidth. The typical behavior of the output signal vs. current should feature a slight tuning of the frequency with respect to the current. At first, it exhibits a $\sim$ 200 MHz red-shift, then reaches to a minimum at $\sim$ 2 mA, and finally shows a slow blue-shift till $\sim$ 3 mA. At even higher currents, we see a broadening of the oscillation signal. As mentioned above, these high-current data are taken at or close to the limit of sustained non-destructive operation.\\ 
A few devices show different behaviours like the one presented in Figure \ref{fig:ideal_meas}(b). Here, the oscillation frequency shows a red-shift when current is less than 2.5mA. In the current range from $\sim$ 2.5 mA to $\sim$ 3 mA, this oscillation frequency has two distinct modes. They are seemingly co-existing with a $\sim$ 300 MHz frequency separation. Finally, at currents above $\sim$ 3 mA, only the higher-frequency mode survives.

In order to compare how different devices were behaving with respect one to another, the extracted frequencies of 7 devices were plotted together in \ref{fig:ideal_meas}(c). Here, we can clearly see the output frequency decreases to a local minimum at around 2 mA and increases at higher currents. Moreover, the device-to-device comparison tells us that there is a frequency variation, taken at the minimum location, of up to $\Delta f \approx 270 MHz$.

\section{Simulations}

\subsection{Methods}
To confirm and further analyze the measurement observations, we performed MuMax3 \cite{vansteenkiste2014design, Leliaert2017} micromagnetic simulations including the random thermal field (at 300 K). We started from an ideal SHNO with a NC width of 120 nm that showed good agreement to measurements. The SHNO is modelled by $512\times512\times1$ cells with a cell size of $3.9\times3.9\times5$~nm$^3$. Moreover, in our simulations we considered a dumping constant $\alpha = 0.022$, a saturation magnetization $M_{sat} = 601.5$~kA/m, an exchange stiffness $A_{ex} = 10^{-12}$~J/m, a gyromagnetic ratio $\gamma_{LL} = 1.85555 \times 10^{11}$~rad/Ts, and a spin-Hall angle $\theta_{SH} = 0.08$. To further analyse how the oscillator is responding to different conditions, various physical parameters were tested. By sweeping one parameter at a time and keeping the others constant, we could isolate the effect of that parameter over the others. In particular, we investigated the influence of a $\pm 5\%$ variation over the magnetic field (and its direction) and the NC width. \par

\begin{figure}[ht!]
\centerline{\includegraphics[width=0.45\textwidth]{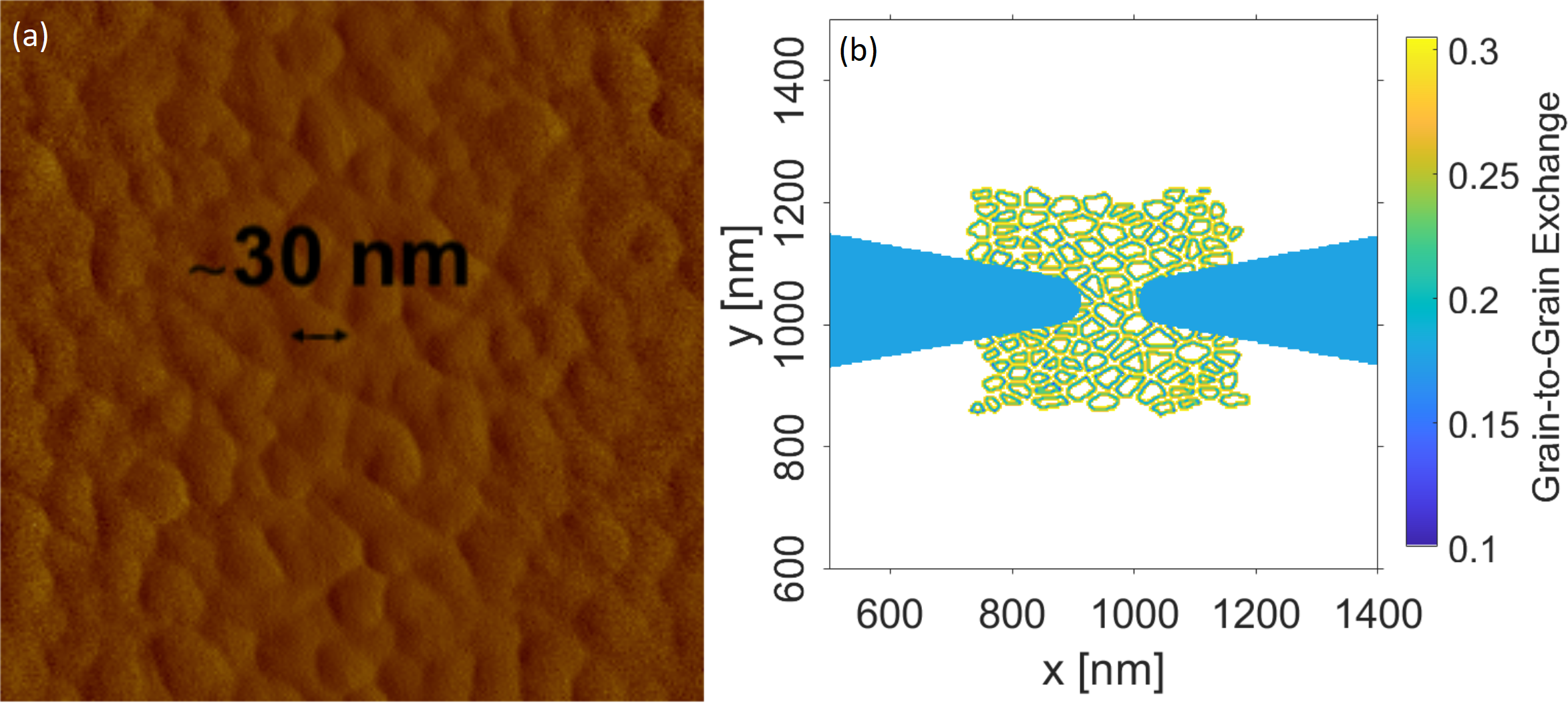}}
\caption{(a) Atomic force microscope (AFM) scan used for the grain extraction process. Estimated average grain size $\sim$ 30 nm. (b) Example of random grain-to-grain exchange coupling distribution. The grain-to-grain exchange coupling was randomly assigned in a range $10\%$ to $30\%$.}
\label{fig:afm}
\end{figure}

In order to include the grains into the simulation, we used a novel technique based on input data obtained from actual samples. First, a real thin-film sample was analysed with an atomic force microscope (AFM), a $5 \times 5 \mu m$ scan revealed a grain size of around 30 nm (Figure \ref{fig:afm}(a), this figure is slightly cropped to avoid AFM artifacts present on the edges). The second step was the digitalisation of the grains, which was completed by using image recognition followed by a Voronoi tessellation, all implemented in Matlab. Finally, the grains were imported as regions into the micromagnetic simulator. In this way, it was possible to change the exchange coupling between each pair of grains (i.e. regions in the MuMax3 input file). (Figure \ref{fig:afm}(b)). Since the oscillation in a SHNO is highly localized, only a limited area, surrounding the constriction, was modelled by the inclusion of a grain map. The outgoing spin wave has a very low amplitude and spin wave (magnon) reflection or scattering has little influence on the final results. The exchange coupling values were drawn from a seeded pseudo random distribution, in this way we were able to change the exchange coupling in the map by simply varying the random seed.\\
The exchange energy was reduced to highlight the effect of the grains. This reduction was limited to a range of 10\% to 50\%, since above 50\% no influence could be noticed. Later on, we further reduced the range to 10-30 \% to better match experimental results.\par
Finally, an investigation of the influence of size of the grains was also carried out. The grain map presented in Figure \ref{fig:afm}(b) better matches the original 30 nm grain dimension, and allows the simulations to describe with more precision the reality of the measurements. During this investigation we decided not to vary the shape of the grains as we wanted to focus on the effects of their positions and dimensions.

\subsection{Results and Discussion}
\begin{figure}[ht!]
\centerline{\includegraphics[width=0.45\textwidth]{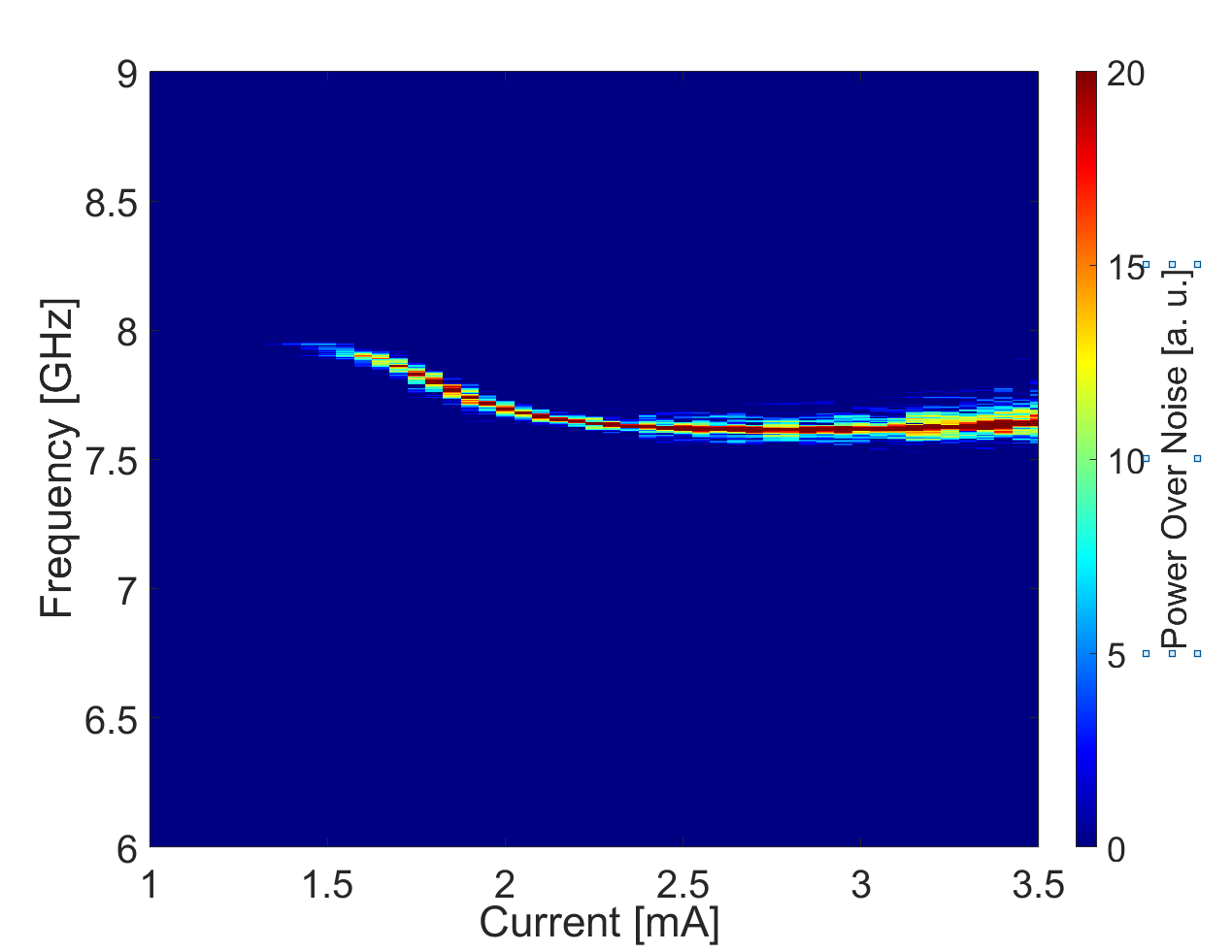}}
\caption{PSD of an ideal 120nm SHNO, simulated for 500 ns at 300 K and 6.5 kOe. This results shows how the output is behaving in a standard way, having a red-shift around 2 mA and a slow blue-shift after 3 mA.}
\label{fig:ideal_sim}
\end{figure}

The simulation of an ideal oscillator shows strong similarities to the baseline measurement (in Figure \ref{fig:ideal_meas}). In  Figure \ref{fig:ideal_sim}, it can be seen how the output is starting at $\sim 1.5$~mA (threshold current), then (at $\sim 2$~mA) a red-shift brings the oscillator towards a frequency minimum $\sim 7.5$~GHz  followed by a slow blue-shift and a broadening of the linewidth (at $\sim 3.3$~mA).\par

In the following section we will first address the effects of the physical parameters (IP-angle and NC width), and then discuss the influence of a random exchange coupling map, based on the grain size and shape extracted from the AFM scan.

\subsubsection{Influence of Field Angle and NC Width}

\begin{figure}[ht!]
\centerline{\includegraphics[width=0.45\textwidth]{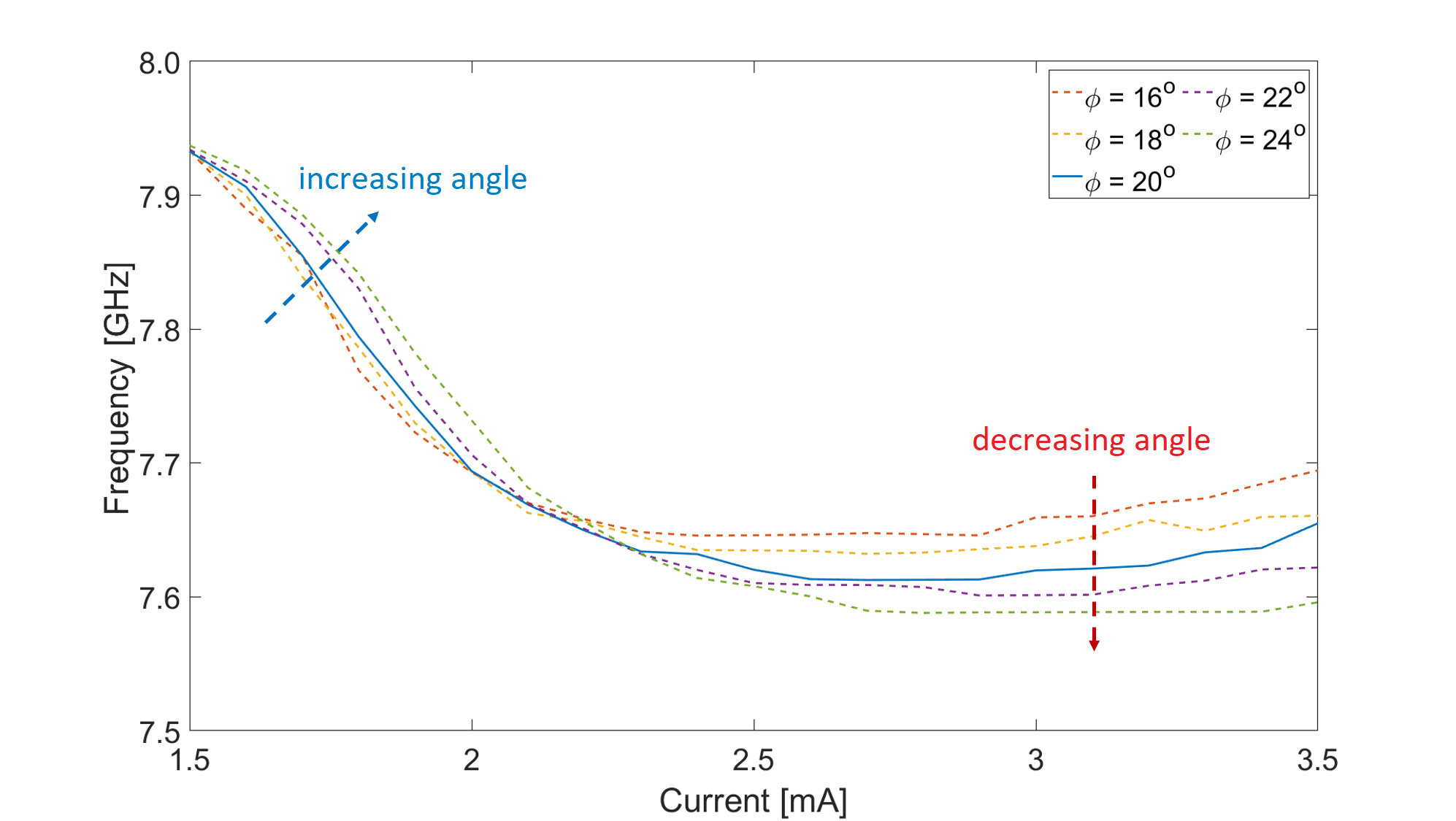}}
\caption{In-plane angle variation effect on SHNO frequency. The IP-angle was varied around the nominal angle, $\phi$ =20\degree}
\label{fig:ip_var}
\end{figure}

First, we focused on the field orientation and the IP-angle by fixing the magnetic field strength (6.5 kOe) and the OOP-angle, $\theta$ = 80\degree. In Figure \ref{fig:ip_var}, we show results for an IP-angle range $\phi$ = 16 - 24\degree. Here, one can observe that the frequencies are independent on the IP angle when the applied current is around 2.1 mA, while at lower and higher currents the frequency range widens. While the IP-angle must be chosen to optimise the anisotropic magnetoresistance (AMR) of the device for maximum output voltage, it does not cause significant instabilities (such as multi-mode behavior) in the typical range discussed here. Any experimentally observed multi-mode behavior should hence have a different physical origin.\par

\begin{figure}[ht!]
\centerline{\includegraphics[width=0.45\textwidth]{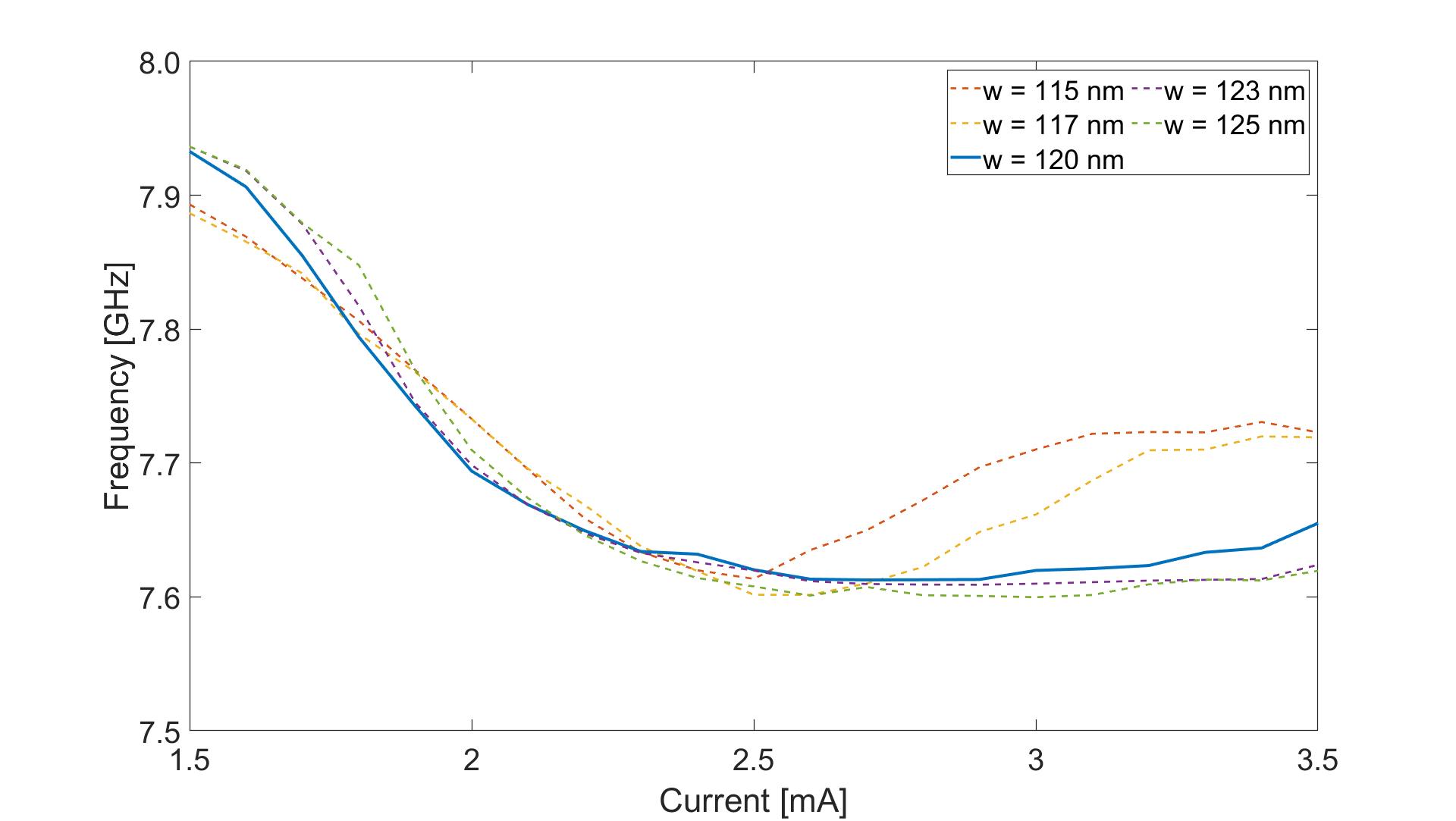}}
\caption{Nano-constriction $\pm$ 5 nm width variation over a nominal value of 120 nm. A pronounced frequency increase at high currents is observed for narrower constriction widths.}
\label{fig:nc_var}
\end{figure}

Then, we study the behaviour of the SHNO at different nano-constriction width. The oscillation frequency is mainly affected in the high current, blue-shifting region, as displayed in Figure \ref{fig:nc_var}. This is consistent with previous work \cite{awad2020} that pointed out that when oscillations propagating from the two edges of the nano-constrinction start to interact, a coherent oscillating volume is formed within the constriction center. As it will be discussed below, the influence of locally reduced exchange coupling will be stronger under these operating conditions, since the oscillator coherence can be directly affected.

\subsubsection{Grain Effects}
\begin{figure}[ht!]
\centerline{\includegraphics[width=0.45\textwidth]{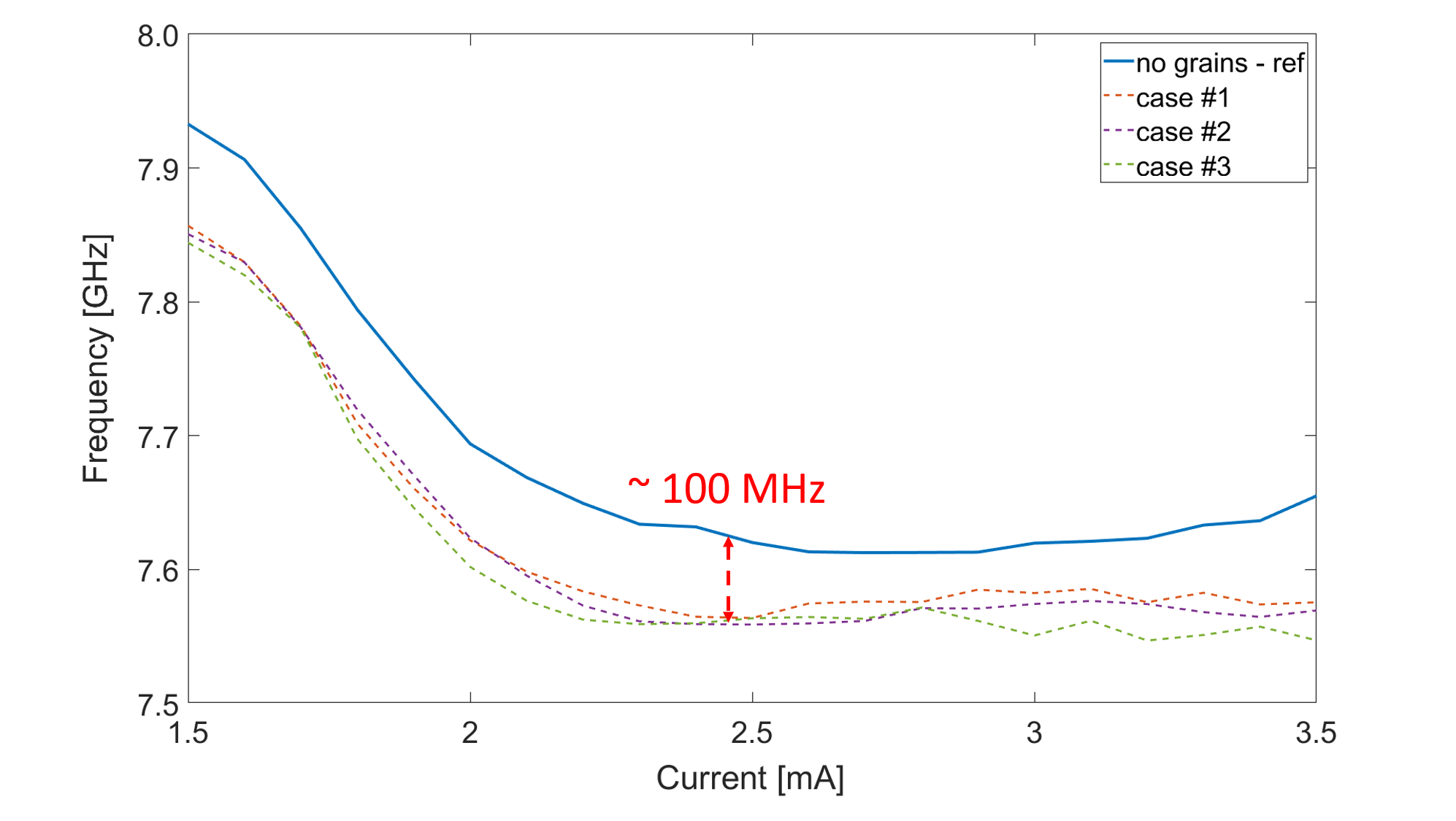}}
\caption{Three different random cases with small grains compared to the ideal device. Here, the exchange coupling was limited to a 10\%--30\%. This reduction causes the oscillatory regime to become unstable after $\sim 3 mA$, almost resembling a double-mode behaviour.}
\label{fig:small_grains_var}
\end{figure}

The main contribution of this work is the introduction of a realistic grain exchange coupling map into the micromagnetic simulations. 
The set of grains was imported in two different ways, a bigger version (with a grain dimension of $\sim 120 nm$) and a smaller version (with a more realistic grain dimension of $\sim 30 nm$). Since we saw a quite similar effect we decided to use the small grains option for all the simulations reported in the following part of this paper.\\
Figure \ref{fig:small_grains_var} reports how the ideal SHNO behaves with small grains under three different conditions, each with a different randomly assigned exchange coupling strength (the coupling was limited to 10\%--30\%). The first noticeable effect of the implementation of the grains results in a device-to-device frequency variability of $\sim 100 MHz$. Another effect can be noticed in the high current range, where the oscillation output of the grain based SHNOs appears much more disturbed than the ideal one and the blue-shift is almost completely gone.

We can see how the frequency shift magnitude in the presence of grains is similar to that of a change in NC width.
This effect can be explained by looking at the situation as follows. Low-exchange grains, within the constriction, can alter the size and shape of the effective oscillator volume. In an extreme limit, the two edges of the constriction could possibly be decoupled, resulting in a pair of two partly phase and frequency incoherent oscillators, instead of one coherent oscillating volume.  \par
\begin{figure}[ht!]
\centerline{\includegraphics[width=0.5\textwidth]{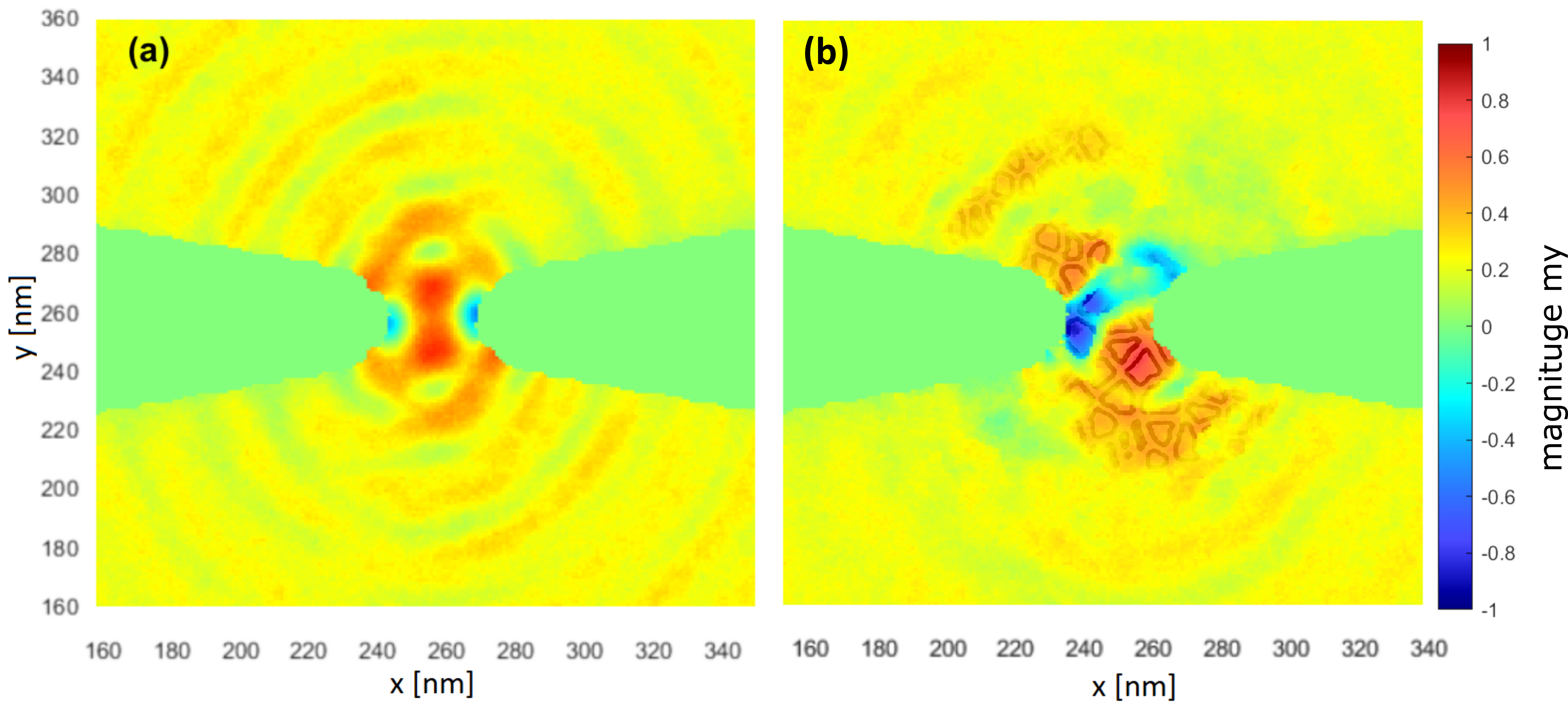}}
\caption{(a) Snap-shot of magnetization vector y-component ($m_y$) at 3.5 mA and 4 K after 7.77 ns of oscillation (ideal case). Oscillatory regime is unperturbed and the oscillation starts from the constriction edges. (b) Magnetization vector y-component ($m_y$) at 3.5 mA and 4 K for case \#3 grain map after 7.77 ns of oscillation. Grain map overlapped to the $m_y$ output of the simulation to highlight how the grains are perturbing the oscillatory regime and de-coupling the two edges.}
\label{fig:my_plot}
\end{figure}

An illustrative way of visualizing the effect of the grains is plotting one the magnetization vector components e.g. $m_y$. A simulation temperature of 4K was used here to reduce the influence of the thermal noise in the plot, the results are in qualitative agreement with 300K data, not shown here. 
In an ideal SHNO (no grain map) (Figure \ref{fig:my_plot}(a)), the oscillatory motion originates at the constriction edges and a coherent oscillator volume is formed in the constriction center. A weak radial spin wave pattern is also discernible, depending on the out of plane angle, the spin wave mode of a NC SHNO will tend to be either bullet like (small angle) or propagating (larger angle). \\
On the other hand, analysing an SHNO with grains, case \#3, the situation looks quite different. In Figure \ref{fig:my_plot}(b), the oscillatory regime looks rather perturbed by the presence of the small grains. In the figure, we overlapped the grain map to the y-component of the magnetization to clarify the effect of the grains.
In particular, we can see how the presence of grains divides the oscillator in two, de-coupling the constriction edges, and thus creating an asymmetry in the spin-wave generation. This configuration is that of two co-existing oscillators, operating without phase and frequency coherence, similar to measured results (Figure \ref{fig:ideal_meas}(b)).

\section{Double-Mode SHNOs}

In our experimental and simulations studies, some devices have exhibited double-mode output characteristics at high currents ($\sim$ 3 mA).\\
In order to confirm the presence of double-mode SHNOs observed during the measurements, long simulations (t = 500 ns) were performed on the small grain devices with a smaller exchange energy reduction rate (10\% - 30 \%). By using this technique, we could see how the oscillatory regime becomes unstable after $\sim 3 mA$, resulting in a double mode behavior.\par
\begin{figure}[ht!]
\centerline{\includegraphics[width=0.5\textwidth]{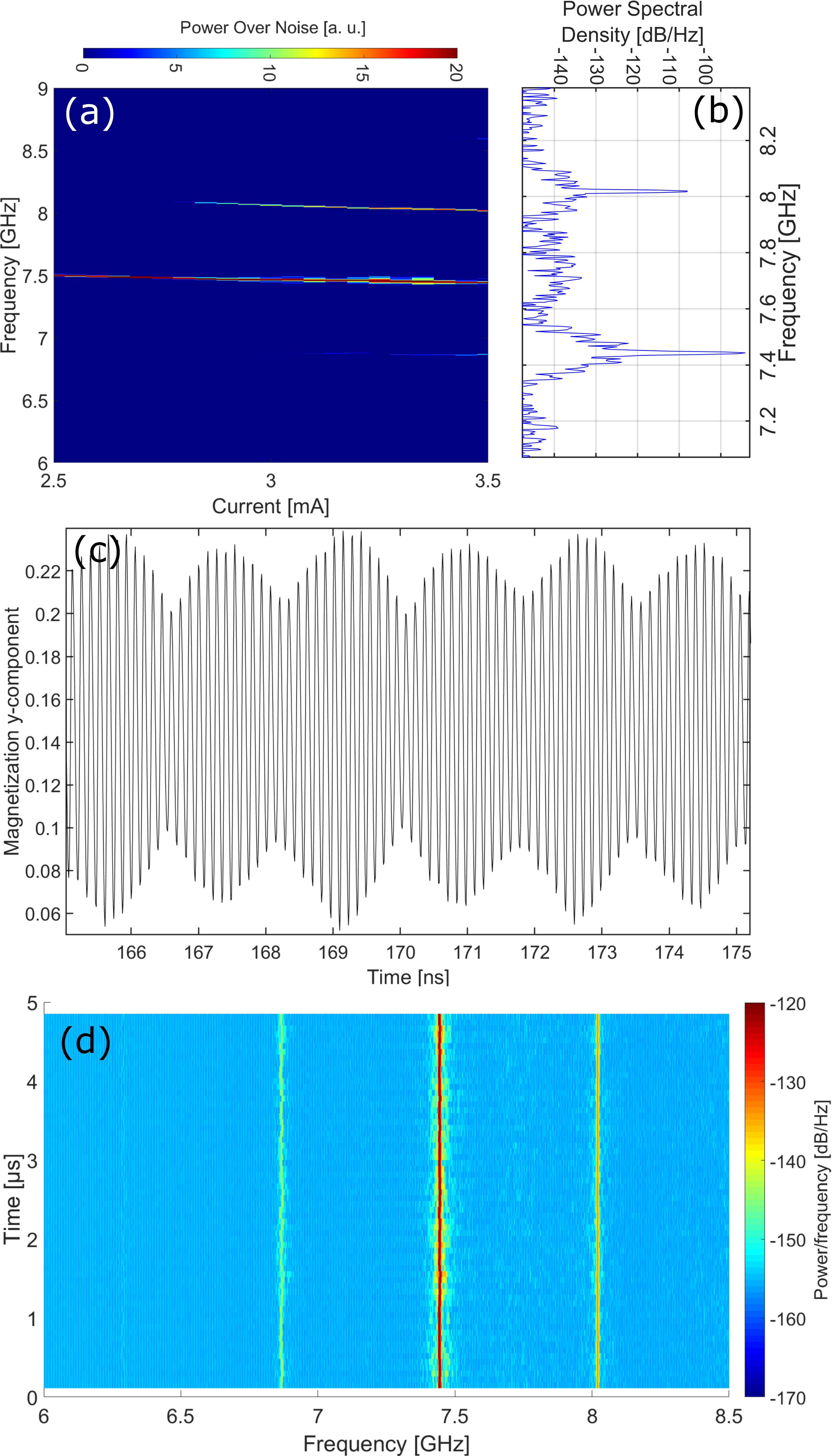}}
\caption{(a) PSD of a simulated device with a double-mode oscillation at 4 K. Between $\sim$ 2.5 mA and $\sim$ 2.8 mA, a single mode is present, after $\sim$ 3.2 mA two distinct modes are co-existing. (b) PSD of $m_y$ at 3.5 mA and 4 K. 
First mode located at $\sim$ 7.5 GHz, second mode at $\sim$ 8 GHz. (c) Time trace of $m_y$ at 3.5 mA and 4 K, beating period $\sim$ 2 ns. (d) Short time Fourier transform (STFT) of $m_y$ that shows coexistence of the two modes.}
\label{fig:double_mode_sim}
\end{figure}

Figure \ref{fig:double_mode_sim}(a) shows an example of these results. In the PSD, we can see how we have a double mode with a $\sim$ 0.5 GHz difference in frequency, which is comparable to the difference found during the measurements (Figure \ref{fig:ideal_meas}(b)). To analyse the evolution of this new mode with respect to current, and to motivate its origin, we followed a 3 step process:
\begin{enumerate}
  \item examine the position of the peaks in the PSD at different currents;
  \item simulate the power behaviour of the device with and without grains at those currents;
  \item compare the different shapes of the oscillating volume of the two cases to highlight the effect of the grains.
\end{enumerate}\par
In Figure \ref{fig:double_mode_sim}(b), one can clearly see the two peaks. The first mode is oscillating at $\sim$ 7.5 GHz, while the second one at $\sim$ 8 GHz. This double-mode behaviour can also be seen in the time domain (Figure \ref{fig:double_mode_sim}(c)). Here, the two peaks can be seen as a  beating of two close frequencies, with a period of $\sim$ 2 ns, corresponding to the 500 MHz separation in the frequency domain. One last proof of the coexistence of the two modes is presented in Figure \ref{fig:double_mode_sim}(d). In the figure, the short time Fourier transform (STFT) is plotted and it shows the 1$^{st}$ and the 2$^{nd}$ modes coexist for the whole 5 $\mu$s simulation time.\par

\begin{figure}[ht!]
\centerline{\includegraphics[width=0.5\textwidth]{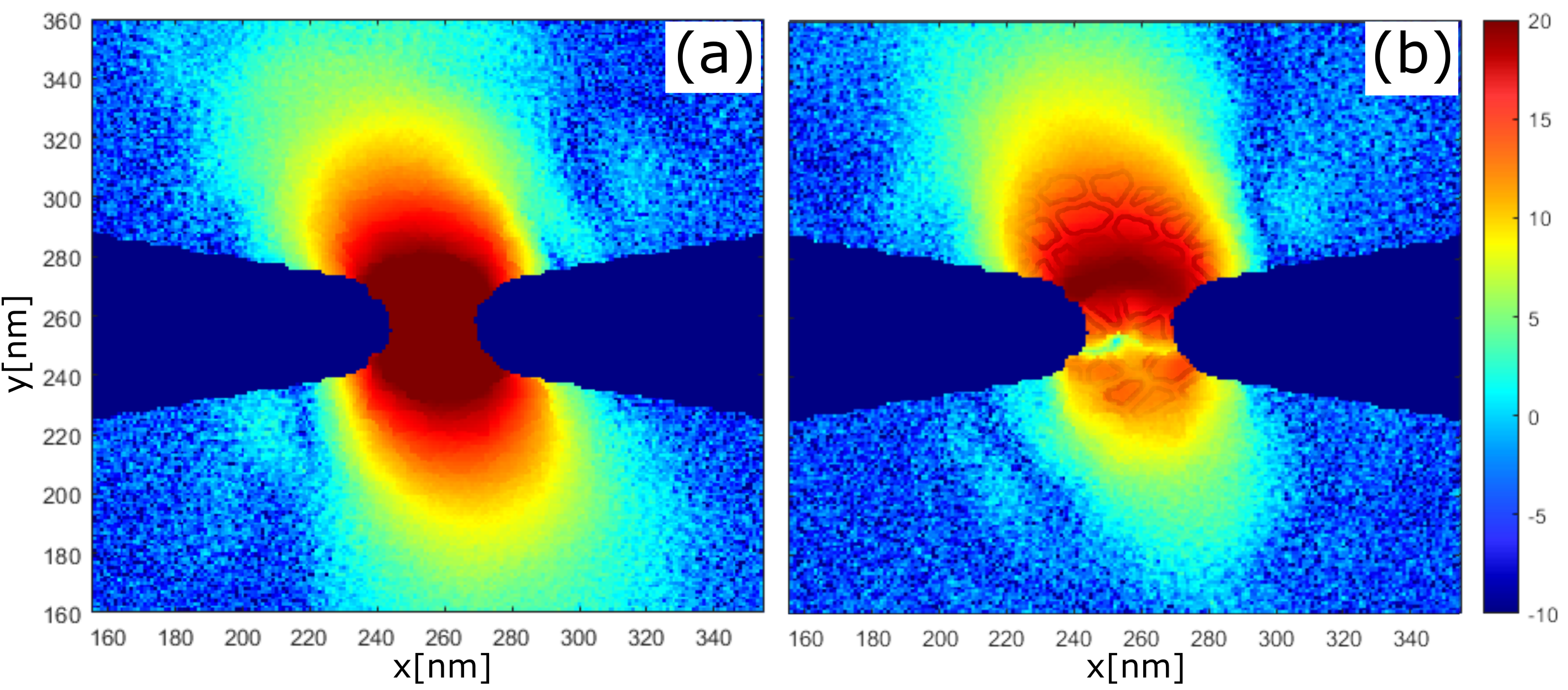}}
\caption{(a) Power plot, and overlapped grain map, of the 1$^{st}$ peak at 3.5 mA and 4 K.  (b) Power plot of the 2$^{nd}$ peak at 3.5 mA and 4 K. The grains are perturbing the oscillating volume confining it in the upper half of constriction.}
\label{fig:pow_plot}
\end{figure}

To explain the double-mode behaviour, the oscillation power landscape was plotted at the respective peak frequencies, $\sim$ 7.5 GHz and $\sim$ 8.0 GHz. These plots are generated from FFTs of the magnetization time evolution in each cell of the MuMax3 simulation domain.
Figure \ref{fig:pow_plot} shows the power plots respectively for the first and for the second peak at 3.5 mA and 4 K. In Figure \ref{fig:pow_plot}(a), a high degree of symmetry is retained in the oscillation power landscape, very similar to plot of results within a uniform domain without local exchange coupling variations. In Figure \ref{fig:pow_plot}(b), however, the oscillating volume is mainly located in the upper half of the constriction, due the reduced local exchange, between pair of grains in the middle of the constriction. \par
Comparing the simulations with the measurements, it is found that the second mode of the simulated device is not strong enough to take over the primary one. In some measured devices the second mode appeared to dominate over the first (lower current mode) mode or at least prohibit their co-existence for very high currents.

\section{\label{sec:conclusion}Conclusions}
To summarise, in this study both experiments and micro-simulations were performed on NC-SHNOs in order to link the presence of physical grains to linewidth broadening and presence of multiple signal frequencies or modes. Statistical measurements, on nominally identical devices, showed a device-to-device frequency variability of $\sim$ 270 MHz. Correspondingly, about $\sim$ 100 MHz device-to-device fluctuation was obtained, by implementing a realistic grain structure in micromagnetic simulations. The grain structure was based on AFM analysis of thin-film samples. The device-to-device difference in frequency was linked to the shape of the oscillating volume. The simulation methodology could also successfully describe a double mode oscillation in the presence of reduced exchange coupling at grain boundaries. Both simulations and measurements showed similar frequency separation $\sim$ 0.5 GHz between the two modes.\\
Further experimental work is suggested, to limit the presence of the grains, either enlarging them or by converting them into amorphous layers.

\section*{Acknowledgment}
This work was supported by the Swedish Research Council (VR) with the project Fundamental Fluctuations in Spintronics [project number 2017-04196].


\begin{thebibliography}{1}

\bibitem{Dieny2020}
Dieny, B., Prejbeanu, I. L., Garello, K., Gambardella, P., Freitas, P., Lehndorff, R., Raberg, W., Ebels, U., Demokritov, S.O., Åkerman, J., Deac, A., Pirro, P., Adelmann, C., Anane, A., Chumak, A.V., Hiroata, A., Mangin, S., Cengiz Onbaşlı, M., d’Aquino, M., Prenat, G., Finocchio, G., Lopez Diaz, L., Chantrell, R., Chubykalo-Fesenko, O., and Deac, A. (2020). Opportunities and challenges for spintronics in the microelectronics industry. Nature Electronics, 3(8), 446-459.

\bibitem{chen2016spin}
Chen, T., Dumas, R. K., Eklund, A., Muduli, P. K., Houshang, A., Awad, A. A., D{\"u}rrenfeld, P., Malm, B. G., Rusu, A., and {\AA}kerman, J. (2016). Spin-torque and spin-Hall nano-oscillators. Proceedings of the IEEE, 104(10), 1919-1945.

\bibitem{malm2019micromagnetic}
Malm, B. G., Eklund, A., and Dvornik, M. (2019). Micromagnetic Modeling of Telegraphic Mode Jumping in Microwave Spin Torque Oscillators. 25th International Conference on Noise and Fluctuations – ICNF, Neuchatel, Switzerland. DOI 10.5075/epfl-ICLAB-ICNF-269282

\bibitem{Zahedinejad2018}
Zahedinejad, M., Mazraati, H., Fulara, H., Yue, J., Jiang, S., Awad, A. A., and Åkerman, J. (2018). CMOS compatible W/CoFeB/MgO spin Hall nano-oscillators with wide frequency tunability. Applied Physics Letters, 112(13), 132404.

\bibitem{Demidov2014}
Demidov, V. E., Urazhdin, S., Zholud, A., Sadovnikov, A. V., and Demokritov, S. O. (2014). Nanoconstriction-based spin-Hall nano-oscillator. Applied Physics Letters, 105(17), 172410.

\bibitem{awad2020}
Awad, A. A., Houshang, A., Zahedinejad, M., Khymyn, R., and Åkerman, J. (2020). Width dependent auto-oscillating properties of constriction based spin Hall nano-oscillators. Applied Physics Letters, 116(23), 232401.

\bibitem{Dvornik2018}
Dvornik, M., Awad, A. A., and Åkerman, J. (2018). Origin of magnetization auto-oscillations in constriction-based spin Hall nano-oscillators. Physical Review Applied, 9(1), 014017.

\bibitem{Mazraati2018}
Mazraati, H., Etesami, S. R., Banuazizi, S. A. H., Chung, S., Houshang, A., Awad, A. A., Dvornik, M., and Åkerman, J. (2018). Auto-oscillating spin-wave modes of constriction-based spin Hall nano-oscillators in weak in-plane fields. Physical Review Applied, 10(5), 054017.

\bibitem{awad2017long}
Awad, A. A., Dürrenfeld, P., Houshang, A., Dvornik, M., Iacocca, E., Dumas, R. K., and {\AA}kerman, J. (2017). Long-range mutual synchronization of spin Hall nano-oscillators. Nature Physics, 13(3), 292-299.

\bibitem{zahedinejad2020two}
Zahedinejad, M., Awad, A. A., Muralidhar, S., Khymyn, R., Fulara, H., Mazraati, H., Dvornik, M., and {\AA}kerman, J. (2020). Two-dimensional mutually synchronized spin Hall nano-oscillator arrays for neuromorphic computing. Nature Nanotechnology, 15(1), 47-52.

\bibitem{Demidov2012}
Demidov, V. E., Urazhdin, S., Ulrichs, H., Tiberkevich, V., Slavin, A., Baither, D., Schmitz, G., and Demokritov, S. O. (2012). Magnetic nano-oscillator driven by pure spin current. Nature materials, 11(12), 1028-1031.

\bibitem{Hirsch1999}
Hirsch, J. E. (1999). Spin hall effect. Physical Review Letters, 83(9), 1834.

\bibitem{zhang2000}
Zhang, S. (2000). Spin Hall effect in the presence of spin diffusion. Physical Review Letters, 85(2), 393.

\bibitem{Kato2004}
Kato, Y. K., Myers, R. C., Gossard, A. C., and Awschalom, D. D. (2004). Observation of the spin Hall effect in semiconductors. Science, 306(5703), 1910-1913.

\bibitem{Saitoh2006}
Saitoh, E., Ueda, M., Miyajima, H., and Tatara, G. (2006). Conversion of spin current into charge current at room temperature: Inverse spin-Hall effect. Applied Physics Letters, 88(18), 182509.

\bibitem{zholud2014microwave}
Zholud, A., and Urazhdin, S. (2014). Microwave generation by spin Hall nanooscillators with nanopatterned spin injector. Applied Physics Letters, 105(11), 112404.

\bibitem{Duan2014}
Duan, Z., Smith, A., Yang, L., Youngblood, B., Lindner, J., Demidov, V. E., Vladislav, E., Demokritov, S. O., and Krivorotov, I. N. (2014). Nanowire spin torque oscillator driven by spin orbit torques. Nature Communications, 5(1), 1-7.

\bibitem{durrenfeld201720}
Dürrenfeld, P., Awad, A. A., Houshang, A., Dumas, R. K., and {\AA}kerman, J. (2017). A 20 nm spin Hall nano-oscillator. Nanoscale, 9(3), 1285-1291.

\bibitem{Ranjbar2014}
Ranjbar, M., Drrenfeld, P., Haidar, M., Iacocca, E., Balinskiy, M., Le, T. Q., Fazlali, M., Houshang, A., Awad, A. A., Dumas, R. K., and {\AA}kerman, J. (2014). CoFeB-based spin Hall nano-oscillators. IEEE Magnetics Letters, 5, 1-4.

\bibitem{Mazraati2016}
Mazraati, H., Chung, S., Houshang, A., Dvornik, M., Piazza, L., Qejvanaj, F., Jiang, S., Le, T. Q., Weissenrieder, J., and Åkerman, J. (2016). Low operational current spin Hall nano-oscillators based on NiFe/W bilayers. Applied Physics Letters, 109(24), 242402.

\bibitem{Haidar2019}
Haidar, M., Awad, A. A., Dvornik, M., Khymyn, R., Houshang, A., and Åkerman, J. (2019). A single layer spin-orbit torque nano-oscillator. Nature Communications, 10(1), 1-6.

\bibitem{dumas2013spin}
Dumas, R. K., Iacocca, E., Bonetti, S., Sani, S. R., Mohseni, S. M., Eklund, A., Persson, J., Heinonen, O., and Åkerman, J. (2013). Spin-wave-mode coexistence on the nanoscale: a consequence of the oersted-field-induced asymmetric energy landscape. Physical Review Letters, 110(25), 257202.

\bibitem{iacocca2015mode}
Iacocca, E., Dürrenfeld, P., Heinonen, O., Åkerman, J., and Dumas, R. K. (2015). Mode-coupling mechanisms in nanocontact spin-torque oscillators. Physical Review B, 91(10), 104405.

\bibitem{Iacocca2014}
Iacocca, E., Heinonen, O., Muduli, P. K., and Åkerman, J. (2014). Generation linewidth of mode-hopping spin torque oscillators. Physical Review B, 89(5), 054402.

\bibitem{Sharma2014}
Sharma, R., Dürrenfeld, P., Iacocca, E., Heinonen, O. G., Åkerman, J., and Muduli, P. K. (2014). Mode-hopping mechanism generating colored noise in a magnetic tunnel junction based spin torque oscillator. Applied Physics Letters, 105(13), 132404.

\bibitem{Heinonen2013}
Heinonen, O. G., Muduli, P. K., Iacocca, E., and Åkerman, J. (2013). Decoherence, mode hopping, and mode coupling in spin torque oscillators. IEEE Transactions on Magnetics, 49(7), 4398-4404.

\bibitem{Zhang2017}
Zhang, S. S. L., Iacocca, E., and Heinonen, O. (2017). Tunable Mode Coupling in Nanocontact Spin-Torque Oscillators. Physical Review Applied, 8(1), 014034.

\bibitem{Eklund2014}
Eklund, A., Bonetti, S., Sani, S. R., Majid Mohseni, S., Persson, J., Chung, S.,  Banuazizi, S. A. H., Iacocca, E., Östling, M., Åkerman,J., and Gunnar Malm, B. (2014). Dependence of the colored frequency noise in spin torque oscillators on current and magnetic field. Applied Physics Letters, 104(9), 092405.

\bibitem{chen2019}
Chen, L., Urazhdin, S., Du, Y. W., and Liu, R. H. (2019). Dynamical mode coupling and coherence in a spin Hall nano-oscillator with perpendicular magnetic anisotropy. Physical Review Applied, 11(6), 064038.

\bibitem{chen2019dynamical}
Chen, L., Zhou, K., Urazhdin, S., Jiang, W., Du, Y. W., and Liu, R. H. (2019). Dynamical mode coexistence and chaos in a nanogap spin Hall nano-oscillator. Physical Review B, 100(10), 104436.

\bibitem{vansteenkiste2014design}
Vansteenkiste, A., Leliaert, J., Dvornik, M., Helsen, M., Garcia-Sanchez, F., and Van Waeyenberge, B. (2014). The design and verification of MuMax3. AIP Advances, 4(10), 107133.

\bibitem{Leliaert2017}
Leliaert, J., Mulkers, J., De Clercq, J., Coene, A., Dvornik, M., and Van Waeyenberge, B. (2017). Adaptively time stepping the stochastic Landau-Lifshitz-Gilbert equation at nonzero temperature: Implementation and validation in MuMax3. Aip Advances, 7(12), 125010.



\end{thebibliography}
\end{document}